\documentclass[showpacs,preprintnumbers,amsmath,amssymb,aps,preprint]{revtex4-1}

\usepackage{graphicx}% Include figure files
\usepackage{dcolumn}% Align table columns on decimal point
\usepackage{bm}% bold math
\usepackage{hyperref}% add hypertext capabilities
\usepackage{lipsum}
\usepackage{times}

\newcommand{\figref}[1]{Fig.~\ref{fig:#1}}
\newcommand{\Figref}[1]{Figure~\ref{fig:#1}}
\newcommand{\citeasnoun}[1]{Ref.~[\onlinecite{#1}]}
\renewcommand{\eqref}[1]{(\ref{eq:#1})}

\newcommand{\eqreftwo}[2]{(\ref{eq:#1}) and (\ref{eq:#2})}

\begin{document}

\title{Enhanced nonlinear frequency conversion and Purcell enhancement
  at exceptional points}

\author{Adi Pick$^{1}$}
\thanks{These authors contributed equally to this work.}
\author{Zin Lin$^{2}$}
\thanks{These authors contributed equally to this work.}
\author{Weiliang Jin$^{3}$}
\author{Alejandro W. Rodriguez$^{3}$}

\affiliation{$^1$Department of Physics, Harvard University, Cambridge, MA 02138}
\affiliation{$^2$John A. Paulson School of Engineering and Applied Sciences, Harvard University, Cambridge, MA 02138}
\affiliation{$^3$Department of Electrical Engineering, Princeton University, Princeton, NJ, 08544}

\date{\today}

\begin{abstract}
  We derive analytical formulas quantifying radiative emission from
  subwavelength emitters embedded in triply resonant nonlinear
  $\chi^{(2)}$ cavities supporting exceptional points (EP) made of
  dark and leaky modes. We show that the up-converted radiation rate
  in such a system can be greatly enhanced---by up to two orders of
  magnitude---compared to typical Purcell factors achievable in
  non-degenerate cavities, for both monochromatic and broadband
  emitters. We provide a proof-of-concept demonstration by studying an
  inverse-designed 2D photonic-crystal slab that supports an EP formed
  out of a Dirac cone at the emission frequency and a phase-matched,
  leaky-mode resonance at the second harmonic frequency.
\end{abstract}

\pacs{Valid PACS appear here} \maketitle

The increase in radiative emission experienced by a subwavelength
particle near a resonant cavity is often characterized by the
well-known Purcell factor~\cite{purcell}. In recent work, we presented
a generalization of Purcell enhancement that applies to situations
involving exceptional points (EP)~\cite{Adi16,ZinEP3}---spectral
singularities in non-Hermitian systems where two or more eigenvectors
and their corresponding complex eigenvalues coalesce, leading to a
non-diagonalizable, defective Hamiltonian. EPs are attended by a slew
of intriguing physical effects~\cite{Moiseyev11, Kato95} and have been
studied in various contexts, including lasers, atomic and molecular
systems~\cite{Berry04, Heiss12}, photonic
crystals~\cite{Zhen15,CerjanEP,ZinEP3}, parity-time symmetric
lattices~\cite{Bender98,Zin11, Feng13,Ruter10, Guo09, Mei10, Hamid12,
  Longhi14, Ge14,Liertzer12, Hodaei14, Feng14, NoriLasing}, and
optomechanical resonators~\cite{Gao15,Harris17,NoriOM}. An important
but little explored property of EPs related to light-matter
interactions is their ability to modify and enhance a related
quantity, the local density of states (LDOS)~\cite{Adi16,ZinEP3}.

In this Letter, we demonstrate that radiative emission at $\omega_e$
from a subwavelength particle, e.g. spontaneous emission or
fluorescence from atoms or radiation from plasmonic antennas, embedded
in a triply resonant nonlinear $\chi^{(2)}$ cavity can be greatly
modified and efficiently up-converted to $2\omega_e$ in the vicinity
of an EP. The efficiency of such a frequency-conversion process
depends strongly on the lifetimes and degree of confinement of the
cavity modes~\cite{Boyd}, which we characterize by deriving a
closed-form, analytical formula for the nonlinear Purcell factor: the
LDOS or emission rate at $2\omega_e$ from a dipole current source
oscillating at $\omega_e$. In particular, we obtain emission bounds
applicable to situations involving both monochromatic and
broad-bandwidth emitters, showing that the nonlinear Purcell factor in
a cavity supporting an EP at $\omega_e$ formed out of dark and leaky
modes can generally be more two orders of magnitude larger than that
of a non-degenerate cavity, depending on the position of the emitter
and on complicated but designable modal selection rules. When combined
with recently demonstrated inverse-designed structures optimized to
enhance nonlinear interactions~\cite{Zin16,ZinEP3}, the proposed EP
enhancements could lead to several orders-of-magnitude larger
luminescence efficiencies.

% under normal, non-degenerate (quasi-)Hermitian systems,i.e.
The key to enhancing the LDOS at an EP is to exploit the intricate
physics arising from the coalescence of dark and leaky (lossy)
resonances. Featuring infinite lifetimes and vanishing decay rates,
dark modes are by definition generally inaccessible to external
coupling. Consequently, an emitter on resonance with a dark mode
cannot radiate unless it is also coupled to a leaky mode.
%in which case the energy radiated through the latter is partially
%enhanced by the former.
Such a shared resonance underlies the monochromatic LDOS enhancements
at EPs described recently in Refs.~\cite{Adi16,ZinEP3}, which showed
that the LDOS at an EP exhibits a narrowed, squared Lorentzian
lineshape whose peak is four times larger than the maximum LDOS at a
non-degenerate resonance. More generally, for an EP of order $n$, the
maximum enhancement factor scales as
$\sqrt{n^3}$~\cite{ZinEP3}. Although such an effect makes it possibile
to enhance monochromatic emission near the EP resonance, the existence
of a sum rule~\cite{barnett1996sum}, which forces the
frequency-intregrated LDOS over the resonance bandwidth to be a
constant, prohibits any enhancement in the case of broadband emitters
(e.g. fluorescent molecules). In this work, we exploit a coupled-mode
theory framework to show that in contrast to the linear LDOS, both the
monochromatic and frequency-integrated radiation rate of a dipolar
emitter in a nonlinear medium can be can be enhanced in the presence
of an EP. We buttress our theoretical predictions with a concrete
physical example: a 2D PhC slab designed to support an EP at
$\omega_e$ and a leaky (phase-matched) resonance at
$2\omega_e$. Furthermore, we consider individual emitters as well as
uniform distributions of incoherent emitters throughout the crystal,
showing that EPs can enhance emission in both cases.

%Technically, in the case of a single, isolated emitter, we refer to
%the LDOS-per-k (also known as the Mutual Density of States~\cite{dos})
%but when integrated over the unit cell, this quantity is known as the
%spectral DOS, which quantifies angular-selective emission from an
%ensemble of emitters~\cite{Marin}, as explained below.

%Finally, we discuss the effects of the EP in scenarios where the
%strength of the nonlinear coupling is so strong that down-conversion
%cannot be neglected, leading to additional enhancements.

\begin{figure}[t!]
\centering \includegraphics[width=0.7\columnwidth]{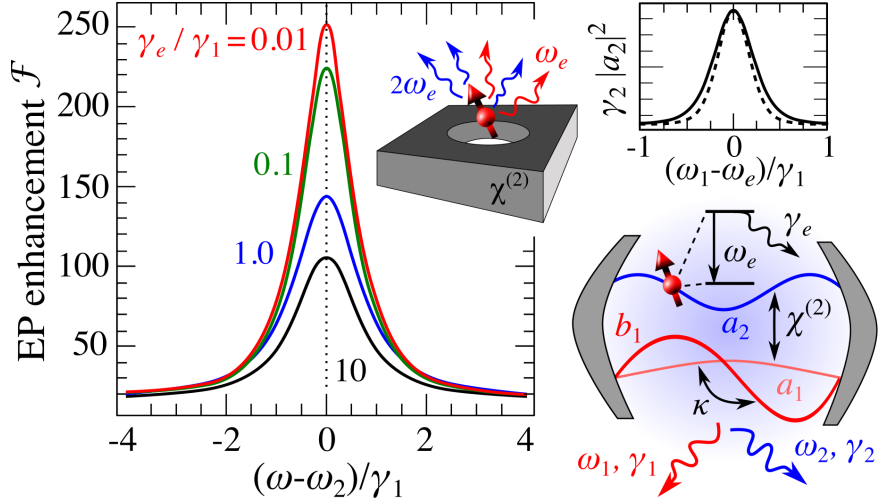}
\caption{\label{fig:fig1} Schematic of a dipole emitter
  $(\omega_e,\gamma_e)$ embedded in a triply resonant $\chi^{(2)}$
  nonlinear cavity supporting an exceptional point (EP). The cavity
  consists of two degenerate resonances $(a_1,b_1)$ at $\omega_1$ that
  are coupled to one another via a linear coupling rate $\kappa$ and
  nonlinearly coupled to a harmonic mode of frequency
  $\omega_2=2\omega_1$ and decay rate $\gamma_2$ by the $\chi^{(2)}$
  process. An EP is formed out of the dark mode $a_1$ and the leaky
  mode $b_1$ of decay rate $\gamma_1$ when $\kappa=\gamma_1/2$. The EP
  leads to enhanced emission at the harmonic frequency, captured in
  the main plot (left) by the EP enhancement factor ${\cal F} =
  \frac{\gamma_1 |a_2^\text{EP}|^2}{\gamma_1|a_2^\text{ND}|^2}$, the
  ratio of the emission rate around $\omega_2$ at the EP to that of a
  non-degenerate (ND) system with a single mode at $\omega_1$ of the
  same effective decay rate $\gamma_1/2$. Here, ${\cal F}$ is plotted
  with respect to the frequency detuning $(\omega-\omega_2)/\gamma_1$
  and for multiple ${\gamma_e \over \gamma_1}$. The upper-right inset
  shows the second-harmonic emission rate $\gamma_2 |a_2|^2$ as a
  function of detuning ${(\omega_e - \omega_1) \over \gamma_1}$ in the
  limit of a monochromatic emitter, $\gamma_e \rightarrow 0$, for both
  EP (solid) and ND (dashed) systems. For convenience, both emission
  rates have been normalized to have the same peak amplitude.}
\end{figure}

\begin{figure}[t!]
\centering \includegraphics[width=0.6\columnwidth]{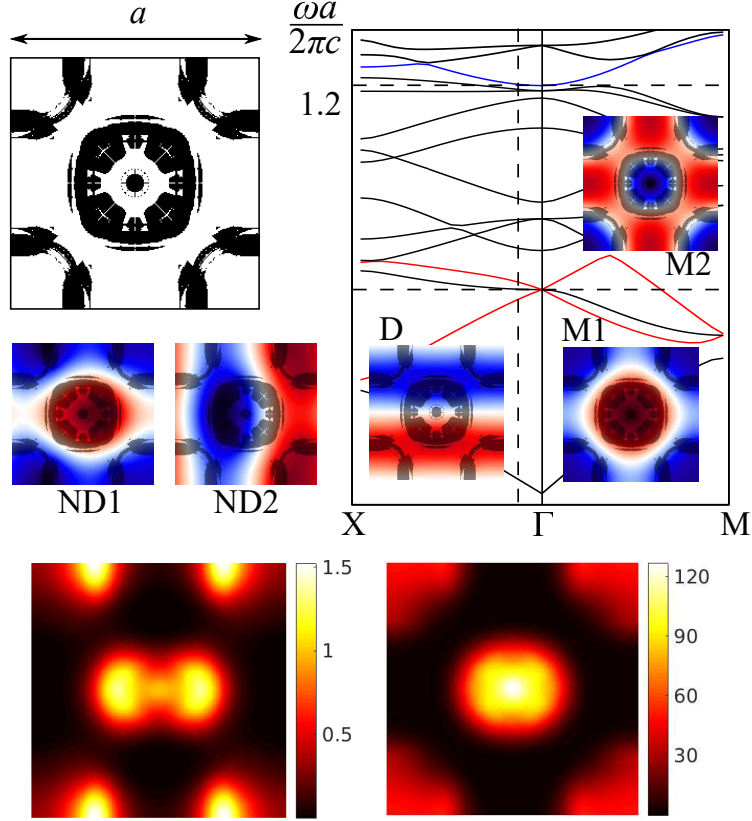}
\caption{\label{fig:fig2} Inverse-designed 2D square PhC (unit
  cell). Dark/white regions represent relative permittivities
  $\epsilon_r=5.5/1$. The corresponding band structure exhibits a
  Dirac cone (red bands) centered at $\omega_1$ and a leaky
  second-harmonic resonance (blue band) around $\omega_2=2\omega_1$. A
  Dirac degeneracy at $\Gamma$ is formed by monopolar (M1) and dipolar
  (D) modes while the second-harmonic resonance is a higher-order
  monopolar mode (M2). Non-hermiticity is introduced by inserting a
  small amount of dielectric loss along the nodal line of the M1 mode,
  allowing realization of an EP near $\omega_1$. Nondegenerate modes
  (ND1, ND2) at $k_xa/2\pi \approx 0.1$ (vertical dashed line in the
  band diagram) are also depicted, illustrating mode mixing between
  the monopole and dipole modes. The lower panels show the spatially
  varying nonlinear LDOS $\int \gamma_2 |a_2|^2
  d\boldsymbol{\mathrm{r}}$, i.e.  the emission rate at $2\omega_e$
  from monochromatic dipole sources oscillating at
  $\omega_e=\omega_1$, in both ND (left) and EP (right) scenarios,
  illustrating a maximum enhancement factor of $\approx 127$ at the
  center of the unit cell. }
\end{figure}

{\it Coupled-mode analysis.---} To understand the impact of EPs on
nonlinear frequency conversion, we consider a generic system involving
a degenerate $(a_1, b_1)$ tuple of dark and leaky modes at $\omega_1$
and a single mode $a_2$ at $\omega_2$. Such a system, shown
schematically in \figref{fig1}, is well described by the following
coupled-mode equations (CME)~\cite{Rodriguez07:OE}:
\begin{align}
{da_1 \over dt} &= i \omega_1 a_1 + i \kappa b_1 - i \omega_1 \left( \beta_1 a_2 a_1^* + \beta_3 a_2 b_1^*\right) + s(t) \label{eq:cme1}\\
{db_1 \over dt} &= i \omega_1 b_1 - \gamma_1 b_1 + i \kappa a_1 - i \omega_1 \left( \beta_2 a_2 b_1^* + \beta_3 a_2 a_1^*\right) \label{eq:cme2}\\
{da_2 \over dt} &= i \omega_2 a_2 - \gamma_2 a_2 -i \omega_1 \left( \beta_1 a_1^2 + \beta_2 b_1^2 + \beta_3 a_1 b_1 \right) \label{eq:cme3}
\end{align}
Mode $a_1$ is dark while $b_1$ and $a_2$ have decay rates $\gamma_1$
and $\gamma_2$, respectively. The two degenerate modes are coupled to
one another via the linear coefficient $\kappa$ and nonlinearly
coupled to $a_2$ by a parametric $\chi^{(2)}$ nonlinear process
characterized by mode-overlap factors
$\beta$s~\cite{Boyd,Rodriguez07:OE}, defined further below in terms of
the linear cavity fields. Solving the CMEs in the absence of
nonlinearities, one finds that for $\kappa \geq \gamma_1/2$, the
frequencies and decay rates of the coupled modes are given by
$\omega_\pm = \omega_1 \pm \sqrt{\kappa^2 - \gamma_1^2/4}$ and
$\gamma_1/2$, respectively, where the latter is independent of
$\kappa$. In particular, the two degenerate modes coalesce at
$\kappa_\text{EP} = \gamma_1/2$, forming an EP at the complex
frequency $\omega_1 - i \gamma_1/2$. In the limit $\kappa \rightarrow
\infty$ of far-apart mode frequencies, one recovers the well-known,
non-degenerate (ND), single-mode description of second-harmonic
generation~\cite{Rodriguez07:OE}, which we compare against when
considering any enhancements arising from the EP~\footnote{Note that
  in principle, coupled-mode theory breaks down in the asymptotic
  limit of infinite coupling $\kappa\to \infty$. However, as
  exemplified in the physical example of \figref{fig2} and as follows
  from the CMEs, similar enhancements are achieved in the more
  practical and adequate situation of far-separated and well-defined
  resonances with $\kappa \gg \gamma_1$.}. Note that in these CMEs,
the term $s(t)$ represents a dipole current source positioned in such
a way so as to exclusively couple to the dark mode, a situation that
is illustrated with a concrete physical example further below. Since
such a term is meant to model a weak emitter (except in the case of
gain media), we primarily focus on the so-called small-signal or
non-depletion regime where one can neglect the nonlinear terms
responsible for down-conversion (e.g. $\beta_1 a_2 a_1^*$).

Before delving further into the nonlinear equations, it is instructive
to briefly review the mechanism of LDOS enhancement in the linear
regime of $\beta=0$. Consider a monochromatic source $s(t) = s_0
e^{i\omega_e t}$ that is on-resonance with the cavity, i.e. $\omega_e
= \omega_1$. Solving the CMEs, one finds that the steady-state mode
amplitudes at the EP are $a_1^\text{EP} = 4 s_0 / \gamma_1$ and
$b_1^\text{EP} = 2 i s_0 / \gamma_1$, whereas in the ND limit of
$\kappa \rightarrow \infty$, $a_1^\text{ND} = b_1^\text{ND} = s_0 /
\gamma_1$. Since the LDOS or radiated power is given by $\gamma_1
|b_1|^2$, it follows that the EP produces an enhancement factor
${\gamma_1 |b_1^\text{EP}|^2 \over \gamma_1 |b_1^\text{ND}|^2 } = 4$,
a result recently derived in~\citeasnoun{Adi16} by a perturbative
expansion of the Green's function based on Jordan eigenvectors but
which also follows from the coupled-mode picture above (see
supplemental materials). We emphasize that such an enhancement can be
realized despite the fact that both the EP and ND resonances exhibit
the same effective decay rate $\gamma_1/2$, indicating that the
enhancement does not arise from an otherwise trivial increase in
resonant lifetimes but rather from a constructive interference of the
two modes, which leads to both narrowing and amplification of the
cavity spectrum~\cite{Adi16}. Unfortunately, such an enhancement
disappears when considering the frequency-integrated emission from a
broadband source, a consequence of a general sum rule (derived from
causality~\cite{barnett1996sum}) which implies that $\int \gamma_1
|b_1^\text{EP}(\omega)|^2~ d\omega = \int \gamma_1
|b_1^\text{ND}(\omega)|^2 ~d\omega$. As we show below, however, such a
sum rule no longer seems to be valid in the case of finite $\beta \neq
0$.

Consider a typical Lorentzian source, $ s(t) = \int_{-\infty}^\infty {
  \sqrt{\gamma_e} \over \gamma_e + i\left(\omega - \omega_e \right)}
e^{i \omega t} ~d\omega$, of frequency $\omega_e$ and decay rate
$\gamma_e$, and whose Fourier amplitude $s(\omega)$ is normalized so
that $\int |s(\omega)|^2~d\omega = \pi$. Solving the CMEs in the
non-depletion regime yields the amplitude $a_2$ of the harmonic mode
as a convolution,
\begin{multline}
a_2(\omega) = { i \omega_1/2 \over i \left( \omega - \omega_2 \right)
  + \gamma_2 } \int_{-\infty}^\infty~dq~\Big[ \beta_1 a_1\left(\omega\right)
  a_1\left(\omega-q\right) \\ +\beta_2 b_1\left(\omega\right)
  b_1\left(\omega-q\right) +\beta_3 a_1\left(\omega\right)
  b_1\left(\omega-q\right) \Big],
\end{multline}
in terms of the mode amplitudes,
\begin{align}
  a_1(\omega) &= \frac{ s(\omega) \sqrt{\gamma_e} \left( \gamma_1 + i \left(\omega - \omega_1 \right)\right)}{\left( \kappa^2 + \left(\omega-\omega_1\right)\left( i \gamma_1 - \omega + \omega_1 \right) \right) \left( \gamma_e + i \left(\omega-\omega_e \right)\right) } \\
  b_1(\omega) &= \frac{ i s(\omega) \sqrt{\gamma_e} \kappa } {\left(
      \kappa^2 + \left(\omega-\omega_1\right)\left( i \gamma_1 -
        \omega + \omega_1 \right) \right) \left( \gamma_e + i
      \left(\omega-\omega_e \right)\right)}.
\end{align}
which can be evaluated to yield closed-form, analytical solutions (see
supplemental materials). In the particular limit of a monochromatic
source with $\gamma_e \ll \gamma_1$, the emission rate at the harmonic
frequency, $\gamma_2 |a_2(\delta)|^2$, or nonlinear LDOS can be
written as:
\begin{widetext}
\begin{align}
%  \gamma_2 |a^\text{EP}_2(\delta)|^2_{\gamma_e \rightarrow 0} =
%  \frac{64 \pi ^2 \zeta |s|^4 \left[16 \beta _1^2 \left(\delta
%        ^2+1\right)^2+4 \beta _3^2 \left(\delta ^2+1\right)+8 \beta _1
%      \left(\beta _2 \left(\delta ^2-1\right)-2 \beta _3 \left(\delta
%          ^3+\delta \right)\right)-4 \beta _2 \beta _3 \delta +\beta
%      _2^2\right]}{\gamma_1^5 \left(4 \delta ^2+1\right)^4 \left(4
%      \delta ^2+\zeta ^2\right)},\label{eq:nldos}
  \gamma_2 |a^\text{EP}_2(\delta)|^2_{\gamma_e \rightarrow 0} &=
  \frac{64 \pi^2 \zeta |s|^4}{\gamma_1^5 \left(4
    \delta^2+1\right)^4\left(4 \delta^2+\zeta^2\right)}
  \Big[16\beta_1^2\left(\delta^2+1\right)^2+4 \beta_3^2
    \left(\delta^2+1\right) \nonumber \\ &\hspace{1.5in} +8 \beta_1 \left(\beta_2
    \left(\delta^2-1\right)-2 \beta_3\left(\delta^3+\delta
    \right)\right)-4 \beta_2 \beta_3 \delta
    +\beta_2^2\Big],\label{eq:nldos}
\end{align}
\end{widetext}
where $\delta = {\omega_1 - \omega_e \over \gamma_1}$ is the
normalized frequency detuning of the emitter from the cavity resonance
and $\zeta = \gamma_2/\gamma_1$. Evidently, the output spectrum
assumes a narrowed and highly non-Lorentzian lineshape, a signature of
the EP. In the opposite limit of a broadband source with $\gamma_e \gg
\gamma_1$, the relevant quantity to consider is the integrated LDOS
near $\omega_2$, given by:
\begin{widetext}
\begin{align}
%\int \gamma_2 |a^\text{EP}_2(\omega)|^2~d\omega \Big|_{\gamma_e \gg \gamma_1}  
%&\approx {\pi^3 |s|^4 \over 4608 \gamma_1^5 (\zeta +1)^3} \left({\gamma_1 \over \gamma_e}\right)^2
%\Bigg[ \beta _1^2 \left(237312 \zeta ^2+638208 \zeta +460800\right)-2 \beta _2 \beta _1 \left(29952 \zeta ^2+89856 \zeta +92160\right) \notag \\
%&+\beta _2^2 \left(6912 \zeta ^2+20736 \zeta +18432\right)+\beta _3^2 \left(29952 \zeta ^2+89856 \zeta +73728\right) \Bigg] \label{eq:nidos},
\int \gamma_2 |a^\text{EP}_2(\omega)|^2~d\omega \Big|_{\gamma_e \gg
  \gamma_1} &\approx {\pi^3 |s|^4 \over 4608 \gamma_1^5 (\zeta +1)^3}
\left({\gamma_1 \over \gamma_e}\right)^2 \Bigg[ \beta _1^2
  \left(237312 \zeta ^2+638208 \zeta +460800\right)\notag
  \\ &-2 \beta _2 \beta _1 \left(29952 \zeta ^2+89856
  \zeta +92160\right) +\beta _2^2 \left(6912 \zeta ^2+20736 \zeta
  +18432\right)\notag \\ &+\beta _3^2 \left(29952 \zeta ^2+89856 \zeta
  +73728\right) \Bigg] \label{eq:nidos},
\end{align}
\end{widetext}
To quantify the impact of these spectral modifications, we compare the
emission rates at the EP against those obtained in the ND scenario,
given by:
\begin{align}
  \gamma_2 |a^\text{ND}_2(\delta)|^2_{\gamma_e \rightarrow 0} =
  \frac{4 \pi ^2 \left(\beta _1+\beta _2+\beta _3\right){}^2 \zeta
    |s|^4}{\gamma_1^5 \left(4 \delta ^2+1\right)^2 \left(4 \delta
      ^2+\zeta ^2\right)}\label{eq:nldos_nd}
\end{align}
\begin{align}
  \int \gamma_2 |a^\text{ND}_2(\omega)|^2~d\omega \Big|_{\gamma_e \gg
    \gamma_1} = \frac{\pi ^3 \left(\beta _1+\beta _2+\beta
      _3\right){}^2 |s|^4 }{\gamma_1^5 (\zeta +1)} \left({\gamma_1
      \over \gamma_e}\right)^2\label{eq:nidos_nd}
\end{align}

\Figref{fig1} shows the nonlinear EP enhancement factor, ${\cal
  F}(\omega,\gamma_e) = {|a_2^\text{EP}(\omega,\gamma_e)|^2 \over
  |a_1^\text{ND}(\omega,\gamma_e)|^2 }$, which is the ratio of the
emission rate around $\omega_2$ at the EP to that in the ND scenario
for the typical situation of an emitter that is resonantly coupled to
the fundamental cavity frequency, i.e. $\omega_e = \omega_1$. In
particular, the figure shows $\mathcal{F}$ as a function of the output
frequency $(\omega-\omega_2)/\gamma_1$ and for multiple values of
$\gamma_e/\gamma_1$ when all of the nonlinear coupling coefficients
except the one pertaining to the dark mode vanish, i.e. $\beta_1 \neq
0,~\beta_2=\beta_3=0$. Such a nonlinear configuration belies one of
the main results of this work, which follows from~\eqref{nldos_nd}
and~\eqref{nidos_nd}: the largest radiation rates and therefore
Purcell enhancements are achieved when the dipole emitter couples
exclusively to the dark mode and when only the latter couples strongly
to the harmonic mode. Evaluating ${\cal F}$ at $\omega=\omega_2$ and
taking the limit of $\gamma_e \to 0$ or equivalently, evaluating the
ratio of~\eqreftwo{nldos}{nldos_nd} in the limit of zero detuning
$\delta=0$, yields a maximum enhancement factor of 256. The
(top-right) inset of \figref{fig1} shows the dependence of the
nonlinear LDOS~\eqref{nldos} with respect to the emitter detuning
$\delta$ in the monochromatic regime $\gamma_e \rightarrow 0$, showing
a slightly narrowed EP spectrum compared to the ND scenario (both
spectra are normalized to have the same peak amplitude for
clarity). Notably, one finds that compared to the linear scenario
discussed above, the nonlinear spectrum undergoes significantly less
narrowing, evidence that the frequency-integrated emission can also be
enhanced. Indeed, focusing in the case of a broadband emitter with
$\gamma_e \gg \gamma_1$, e.g. a fluorescent molecule~\cite{Marin}, and
taking the ratio of~\eqreftwo{nidos}{nidos_nd}, one finds that the
frequency-integrated emission can be enhanced by a factor of 100.

The aformentioned LDOS enhancements at the EP can be understood
intuitively from a recently derived sum rule~\cite{barnett1996sum}.
In the linear regime, causality demands that when two non-degenerate
resonances of equal bandwidths $\gamma$ merge to form an EP, the
resulting LDOS spectrum becomes a squared Lorentzian~\footnote{The
  optical response of a system near a doubly degenerate leaky
  resonance can be approximated by a second-order, complex pole~[ref]}
and obeys the sum rule~\cite{barnett1996sum}, $\sum_{i} \int
{s_\text{i}^2 \gamma \over \delta^2 + \gamma^2}~d\delta = \int { 2
  s^2_\text{EP} \gamma^3 \over \left( \delta^2 +
  \gamma^2\right)^2}~d\delta$, where $s^2_\text{EP} = s_\text{1}^2 +
s_\text{2}^2$ and $s_\text{1,2}$ denote the coupling strengths of a
dipole source which generally couples to both modes.  It follows from
the sum rule that at an EP and in the special case of identical
coupling strengths, $s_\text{1} = s_\text{2} = s_\text{ND}$, the mode
volume of the cavity resonance decreases (and hence the coupling rate
of an emitter increases) so that $s_\text{ND}=s_\text{EP}/\sqrt{2}$,
but only at the expense of an effectively narrower cavity bandwidth.
Such a multi-modal interference phenomenon also leads to an effective
increase in the nonlinear coupling coefficient, with $\beta_\text{ND}
= 0.5 \beta_\text{EP}$ (see supplemental materials). Both effects
combine to increase nonlinear emission by two orders of magnitude.

{\it Proof-of-concept demonstration.---} One way to realize an EP of
coalescent dark and leaky modes is by exploiting Dirac
cones~\cite{Zhen15}, which are linear conical dispersions in the band
structure of PhCs formed out of the degeneracy of modes belonging to
different symmetry representations. Here, we employ recently developed
inverse-design techniques~\cite{ZinEP3} to design a proof-of-concept
PhC exhibiting a Dirac cone at $\omega_1=\omega_e$ and a leaky mode at
$2\omega_1$, both realized at the $\Gamma$ point ($\mathbf{k}=0$) of
the crystal. \Figref{fig2}(a) shows a schematic of the PhC unit cell
of size $a \times a$, with the Dirac point at $\omega_1$ formed by an
accidental degeneracy of monopolar (M1) and a dipolar (D) modes, while
the second-harmonic resonance consists of a monopolar, higher-order
field (M2). Here, black/white regions denote relative dielectric
$\epsilon_r = 5.5$ and vacuum $\epsilon_r = 1$ permittivities,
respectively. We introduce non-Hermiticity to the system by adding a
small amount of absorption ($\mathrm{Im}[\epsilon_r]\neq 0$) along the
nodal line of M1, which renders the other two modes (D and M2) leaky
while keeping M1 dark. By analogy, we identify M1 as $a_1$, D as
$b_1$, and M2 as $a_2$. Given the mode profiles (insets), we employ
perturbation theory~\cite{phcbook} to obtain the correspoding decay
rates, $\gamma_\text{D}/\omega_1 \approx 10^{-4}$ and
$\gamma_\text{M2} \approx 2 \gamma_\text{D}$. As described
in~\citeasnoun{ZinEP3}, the band structure of the PhC in the vicinity
of $\omega_1$ can be described by an effective $2 \times 2$
Hamiltonian~\cite{Zhen15},
%\begin{align*}
$
\begin{pmatrix}
\omega_1 & v_g k \\
v_g k & \omega_1 + i \gamma_\text{D}
\end{pmatrix}$,
%\end{align*}
where $k$ denotes the Bloch wave number and $v_g$ is the group
velocity (the slope of the conical dispersion). Such a system exhibits
an EP at $k_\text{EP} = \gamma_\text{D}/2v_g$. Assuming that
dielectric regions possess a non-zero second-order susceptibility
$\chi^{(2)}$, the nonlinear overlap factors are given by:
\[\beta = \frac{\int
  \chi^{(2)} E_\text{M2}^* E_i E_j~d\mathbf{r}}{\sqrt{\int \epsilon_r
    |E_\text{M2}|^2~d\mathbf{r}}\sqrt{\int \epsilon_r
    |E_i|^2~d\mathbf{r}}\sqrt{\int \epsilon_r |E_j|^2~d\mathbf{r}}},\]
with $i,j \in \{\text{M1, D}\}$ and $E_{\{\text{M1,M2,D}\}}$ denoting
the electric fields of the M1, M2 and D modes.

Our choice of mode symmetries guarantees nearly optimal nonlinear
coupling coefficients for enhancing ${\cal F}$. In particular, we find
that in this structure, $\beta_1 \approx 0.07
(\chi^{(2)}/\sqrt{\lambda_1^2})$, $\beta_2/\beta_1 \approx 0.04$, and
$\beta_3/\beta_1 \approx 0$. Moreover, by construction a dipole
emitter located at the center of the unit cell couples solely to the
dark mode (M1). In the coupled-mode theory framework, the strength of
the internal dipole current coupling to M1 is given by
$s(\mathbf{r}_0,t) = s(t) E_\text{M1}\left(\mathbf{r}_0\right)/2 \int
\epsilon_r |E_\text{M1}|^2~d\mathbf{r}$, where $\mathbf{r}_0$ is at
the center of the unit cell. Note that technically, what one computes
at $k_\mathrm{EP}$ is the LDOS-per-k or so-called mutual
DOS~\cite{dos}, corresponding to emission from an array of coherent,
dipole emitters periodically placed at the center of each unit
cell. Hence, angular emission is channeled into the EP modes at
$k_\text{EP}$ and up-converted into the corresponding phase-matched
second harmonic mode at $2 k_\text{EP}$. For sufficiently small
$\gamma_\text{D} \propto k_\text{EP} \approx 0$, the phase-matching
condition ($k_2 \approx 2 k_\text{EP}$) can be enforced via
perturbative fine-tuning of the dielectric structure with the aid of
well-known experimental techniques, e.g. thermal, mechanical, or
electro-optic post-fabrication mechanisms~\cite{tuning1,tuning2}. Note
that our system represents a proof of concept, but that it is
straightforward (though computationally intensive) to consider
extensions to 3D slab geometries, in which case the non-Hermiticity
could stem from radiative rather than dielectric losses.

%For comparison (below), we

%In particular, we find that $s_\text{ND1}(\mathbf{r}_0)\approx 1.15
%s_\text{ND2}(\mathbf{r}_0)$, whereas $s_\text{ND1}^2(\mathbf{r}_0) +
%s_\text{ND2}^2(\mathbf{r}_0) \approx s^2(\mathbf{r}_0)$ as is
%consistent with the sum rules (discussed above).  Furthermore,
%nonlinear overlap coefficients are found to be $\beta_\text{ND1}
%\approx 0.62 \beta_1$ and $\beta_\text{ND2} \approx 0.41 \beta_1$

%\emph{Discussion.---} 
\Figref{fig2} also shows the mode profile of two ND modes at $k_x a/2
\pi \approx 0.1$ (vertical dashed line in the band diagram) which
merge to form the EP in the limit as $k_x \rightarrow
k_\text{EP}$. Note that in this example, the two modes (denoted as ND1
and ND2) exhibit equal decay rates but slightly different coupling
strengths, $s_\text{ND1} \approx 0.76 s_\text{EP}$ and $s_\text{ND2}
\approx 0.64 s_\text{EP}$, and nonlinear coefficients,
$\beta_\text{ND1} \approx 0.62\beta_{\text{EP}}$ and $\beta_\text{ND2}
\approx 0.41 \beta_{\text{EP}}$. Comparing against ND1, i.e. the more
localized of the two ND resonances, we find that the EP leads to
monochromatic and frequency-integrated enhancement factors of 127 and
32, respectively. Another important quantity characterizing SE from
PhCs is the spectral density of states~\cite{dos} (SDOS), obtained by
integrating the LDOS-per-k over the entire unit cell. The SDOS
quantifies large-area emission from an incoherent ensemble of dipole
emitters uniformly distributed throughout the PhC, and is relevant to
wide-area fluorescence and lasing~\cite{Marin}. In the above
formulation, the nonlinear SDOS is given by $\int \gamma_2
|a_2(\mathbf{r},\omega)|^2~d\mathbf{r}$, where the spatial dependence
in the mode amplitude $a_2$ comes from the variation of the dipole
coupling in space. More precisely, the coupling of a source at
$\mathbf{r}$ to the fundamental modes is represented by two separate
source terms, $s_{a}(\mathbf{r}) = { E_\text{M1}(\mathbf{r}) \over 2
  \int \epsilon_r E_\text{M1}(\mathbf{r})~d\mathbf{r}}$ and
$s_{b}(\mathbf{r}) = { E_\text{D}(\mathbf{r}) \over 2 \int \epsilon_r
  E_\text{D}(\mathbf{r})~d\mathbf{r}}$, in the equations for $a_1$ and
$b_1$, respectively (see supplemental materials). In contrast,
\eqref{cme1}--\eqref{cme3} capture only the optimal situation in which
$\mathbf{r} = \mathbf{r}_0$ and hence $s_{b}(\mathbf{r}_0) = 0$. Given
the concrete design above, one can compute the spatially varying
coupling coefficients for both the EP and ND scenarios, which are
plotted in the lower panel of~\figref{fig2}. For broadband emitters
with linewidths $\gamma_e \gg \gamma_1$, the quantity of interest is
the frequency-integrated SDOS, in which case the enhancement ratio,
$\frac{\int \int |a_2^\text{EP}|^2 d\omega d\mathbf{r}}{\int \int
  |a_2^\text{ND}|^2 d\omega d\mathbf{r}}$, which compares
second-harmonic emission rates into the selective angular channel
specified by $\mathbf{k}$ (near-normal incidence), is 15.

\emph{Concluding remarks.---} To summarize, we have shown that the
efficiency of nonlinear frequency conversion processes can be greatly
enhanced in cavities featuring EPs. Our derived bounds on the possible
nonlinear Purcell factors achievable in EP systems show that the
degree of enhancement depends on complicated but tunable modal
selection rules and are optimal when the emission sources couple
primarily to dark modes. In combination with recently demonstrated
inverse-designed structures optimized to enhance nonlinear
overlaps~\cite{Zin16}, the proposed EP enhancements could lead to
orders-of-magnitude larger nonlinear interactions and emission
efficiencies.  While luminescence enhancements at EPs in linear media
are nullified in the case of broadband emitters, nonlinear Purcell
factors can be enhanced by two orders of magnitudes even when the
emission bandwidth is much larger than the cavity bandwidth.  Although
we illustrated these ideas by examining a simple proof-of-concept 2d
PhC design, these predictions could also be tested in a wide variety
of structures, including highly nonlinear mid-infrared quantum
wells~\cite{midIR} or microwave super-conducting qubit~\cite{qubit}
platforms. Finally, we expect that similar or even potentially larger
enhancements can arise in systems supporting higher-order exceptional
points~\cite{ZinEP3} or other nonlinear processes, e.g. third-harmonic
generation, four-wave mixing, and two-photon down-conversion, with
potential applications to quantum information science.

\section{Supplement}
  We provide a more detailed analysis of the coupled-mode equations in
  the main text, which describe two optical resonances (a dark and
  leaky mode) linearly coupled to one another to form an exceptional
  point (EP), and nonlinearly coupled to a second-harmonic mode via a
  Pockels $\chi^{(2)}$ medium. We show that relative to non-degenerate
  scenarios, radiative emission at the up-converted frequency from a
  source which couples only to the dark mode is enhanced due to an
  effective reduction in the mode volume of the cavity at the emission
  wavelength, which manifests as an increase in the local density of
  states and nonlinear overlap coefficients governing second-harmonic
  generation. Finally, we provide a full formula for the up-converted
  (second-harmonic) emission rate of a dipole oscillating at the EP
  frequency.

\subsection{Spectral narrowing and mode reduction at an exceptional point}

In Ref.~\cite{Adi16}, we showed via direct manipulation of Maxwell's
equations that an optical cavity supporting an EP will exhibit an
amplified but narrowed (squared Lorentzian) spectrum. We also
exploited a well-known sum-rule which states that the
frequency-integrated density of states must be
conserved~\cite{barnett1996sum} in order to argue that the modified
LDOS lineshape necessarily sets an upper bound of four on the peak
enhancement.  Below, we show that both the spectral enhancement and
sum rule also follow from the CMEs above.

Consider the linearized, coupled-mode equations ($\beta = 0$):
\begin{align}
{da_1 \over dt} &= i \omega_1 a_1 + i \kappa b_1 + s(t) \\
{db_1 \over dt} &= (i \omega_1 - \gamma_1) b_1 + i \kappa a_1. 
\end{align}
The LDOS spectrum of the system is given by:
\begin{align}
\gamma_1 |b_1(\omega)|^2 = \frac{\gamma_1 \kappa^2 |s(\omega)|^2}{\left[\kappa^2 - (\omega - \omega_1)^2\right]^2 + \gamma_1^2 (\omega-\omega_1)^2}
\end{align} 
It follows that at the EP, $\kappa = \gamma_1/2$, the spectrum becomes
a pure, squared Lorentzian,
\begin{align}
\gamma_1 |b_1^\text{EP}(\omega)|^2 = { 2 |s(\omega)|^2 \left({\gamma_1 \over 2}\right)^3 \over \left[ \left({\gamma_1 \over 2}\right)^2 + (\omega - \omega_1)\right]^2},
\label{eq:EP0}
\end{align}
centered around $\omega_1$ and with bandwidth $\gamma_1/2$. Taking the
strong-coupling limit $\kappa \gg \gamma_1$ of two coupled but highly
non-degenerate (ND) resonances, one finds:
\begin{multline}
\gamma_1 |b_1^\text{ND}(\omega)|^2 = {{\gamma_1 \over 4}{|s(\omega)|^2} \over \left({\gamma_1 \over 2} \right)^2 + \left[\omega - (\omega_1 + \kappa)\right]^2} + {{\gamma_1 \over 4}{|s(\omega)|^2} \over \left({\gamma_1 \over 2} \right)^2 + \left[\omega - (\omega_1 - \kappa)\right]^2} - {\gamma_1 |s(\omega)|^2 \over 2 \kappa^2} + \mathcal{O}\Big[{\gamma_1 \over \kappa}\Big]^3
\label{eq:ND0}
\end{multline}
Hence, in the $\kappa \to \infty$ limit, the spectrum becomes a sum of
identical Lorentzians centered around the eigenfrequencies $\omega_1
\pm \kappa$ and with bandwidths $\gamma_1/2$. Notably, their
individual peak amplitudes are exactly four times smaller than that of
the squared Lorentzian:
\begin{align} 
{ \gamma_1 |b_1^\text{EP}(\omega=\omega_1)|^2 \over
    \gamma_1 |b_1^\text{ND}(\omega = \omega_1 \pm \kappa)|^2 } =
  4.  
\end{align} 
Essentially, one could argue that at an EP, the mode volume of the
cavity resonance experiences an effective reduction of $\sqrt{2}$ but
only at the expense of a narrower bandwidth. In particular,
integrating \eqreftwo{EP0}{ND0}, one finds thatfg

\begin{align}
  \int \gamma_1 |b_1^\text{EP}(\omega)|^2~d\omega = \int \gamma_1
  |b_1^\text{ND}(\omega)|^2~d\omega,
\end{align}
in agreement with the aforementioned sum rule~\cite{barnett1996sum}.

\subsection{Nonlinear-overlap enhancement}

In the main text, we showed and argued that the effectively smaller
mode volume associated with an EP also leadsfg to a two-fold increase
in the nonlinear overlap coefficients. Here, we show explicitly how
such an increase manifests in the CMEs.

The Hamiltonian corresponding to the linear, coupled-mode system above
is given by:
\begin{align}
\mathcal{H}=\begin{pmatrix}
  \omega_1 & \kappa \\
  \kappa   & \omega_1 - i \gamma_1
\end{pmatrix}.
\end{align}
For $\kappa \neq \gamma_1/2$, $\mathcal{H}$ can be diagonalized such
that the mode amplitudes $\vec{a}_1=(a_1,b_1)$ can be transformed into
the diagonal basis $\vec{a}'_1=(a'_1,b'_1)$ by a linear, unitary
transformation matrix $S$, such that $\vec{a}_1 = S \vec{a}'_1$. In
the strong-coupling limit $\kappa \rightarrow \infty$ of
ND resonances, the transformation matrix is
$
  S =
\begin{pmatrix}
-{1 \over \sqrt{2}} & {1 \over \sqrt{2}} \\
{1 \over \sqrt{2}} & {1 \over \sqrt{2}} 
\end{pmatrix}
$
Writing the amplitude of the second-harmonic mode in the non-depletion
limit,
\begin{align*}
a_2(\omega) = { -i \omega_1 \left( \beta_1 a_1^2 + \beta_2 b_1^2 + \beta_3 a_1 b_1 \right) \over \gamma_2 + i(\omega - \omega_2) },
\end{align*}
in terms of the ND resonance (i.e. taking $\vec{a}_1 \to S
\vec{a}'_1$), one finds that the amplitude in the ND limit $\kappa \to
\infty$ is given by:
\begin{align}
  a_2(\omega) = { -i \omega_1 \over \gamma_2 + i(\omega - \omega_2) }
  {\left( \beta_1 + \beta_2 + \beta_3 \right) \over 2}
  \left(a'_1\right)^2.
\end{align}
Thus, the nonlinear overlap coefficient of the ND system is related to
the corresponding overlap factors of the system at the EP by the
relation:
\begin{align}
\beta^\text{ND} =  {\left( \beta_1 + \beta_2 + \beta_3 \right) \over 2} .
\end{align}

\subsection{Nonlinear EP enhancement formula}

The nonlinear CMEs describing emission from a dipolar source embedded
in the triply resonant cavity above are (in the non-depletion
regime):
\begin{align}
  {da_1 \over dt} &= i \omega_1 a_1 + i \kappa b_1 +
  s_a(t) \label{eq:cme1}\\ {db_1 \over dt} &= (i \omega_1 - \gamma_1)
  b_1 + i \kappa a_1 + s_b(t) \label{eq:cme2}\\ {da_2 \over dt} &= i
  \omega_2 a_2 - \gamma_2 a_2 -i \omega_1 \left( \beta_1 a_1^2 +
  \beta_2 b_1^2 + \beta_3 a_1 b_1 \right) \label{eq:cme3}
\end{align}
Here, we assume that the two cavity resonances are frequency-matched
for second-harmonic generation, so that $\omega_2 = 2
\omega_1$. Assuming a Lorentzian dipole source located at some
position $\vec{r}$, the coupling amplitudes in \eqreftwo{cme1}{cme2}
are $s_{a/b}(t) = \left(\frac{E_{a/b}(\vec{r})}{2 \int \epsilon_r
  |\vec{E}_{a/b}|^2 d\vec{r}}\right) \int_{-\infty}^\infty {
  \sqrt{\gamma_e} \over \gamma_e + i\left(\omega - \omega_e \right)}
e^{i \omega t} ~d\omega$. To obtain an explicit expression for
$a_2(\omega)$, it suffices to Fourier transform \eqref{cme3}, in which
case one finds that the amplitude at the second harmonic depends on a
convolution of the linear modes at $\omega_1$. Focusing on the EP
scenario ($\kappa=\gamma_1/2$) and defining $\delta =
\omega-2\omega_1$, one obtains: \small
\begin{widetext}
\begin{align}
&a^\text{EP}_2(\omega)=-8 \pi \gamma _e \Bigg\{ i \beta _3 \Big[ 2 s_a
    s_b \left(\gamma _1^2 \delta+i \gamma _1^3+\gamma _1 \delta
    \left(4 \gamma _e+i \left(-4 \omega _e+5 \omega -6 \omega
    _1\right)\right) +2 \delta^2 \left(i \gamma _e+\omega _e-\omega
    +\omega _1\right)\right) \notag \\ &+\gamma _1 s_a^2 \left(2
    \gamma _1+i \delta\right) \left(2 \gamma _1+2 \gamma _e+i \left(-2
    \omega _e+3 \omega -4 \omega _1\right)\right) +\gamma _1 \delta
    s_b^2 \left(2 i \gamma _1+2 i \gamma _e+2 \omega _e-3 \omega +4
    \omega _1\right) \Big] \notag \\ &+\beta _1 \Big[ 2 \gamma _1 s_a
    s_b \left(2 \gamma _1+i \delta\right) \left(2 i \gamma _1+2 i
    \gamma _e+2 \omega _e-3 \omega +4 \omega _1\right) \notag
    \\ &+s_a^2 \left(8 \gamma _1^3 +\gamma _1^2 \left(10 \gamma _e+i
    \left(-10 \omega _e+21 \omega -32 \omega _1\right)\right) +4
    \gamma _1 \delta \left(3 i \gamma _e+3 \omega _e-4 \omega +5
    \omega _1\right)\right.\notag \\ &\left.-4 \delta^2\left(\gamma
    _e+i \left(-\omega _e+\omega -\omega _1\right)\right)\right)
    +\gamma _1^2 s_b^2 \left(-2 \gamma _1-2 \gamma _e+2 i \omega _e-3
    i \omega +4 i \omega _1\right) \Big] \notag \\ &-\beta_2 \Big[2
    \gamma _1 \delta s_a s_b \left(2 \gamma _1+2 \gamma _e+i \left(-2
    \omega _e+3 \omega -4 \omega _1\right)\right) +\gamma _1^2 s_a^2
    \left(2 \gamma _1+2 \gamma _e+i \left(-2 \omega _e+3 \omega -4
    \omega _1\right)\right) \notag \\ &+s_b^2 \left(\gamma _1^2
    \left(-2 \gamma _e+2 i \omega _e-i \omega \right) +4 \gamma _1
    \delta \left(-i \gamma _e-\omega _e+\omega -\omega _1\right)+4
    \delta^2 \left(\gamma _e+i \left(-\omega _e+\omega -\omega
    _1\right)\right)\right) \Big] \Bigg\} \notag \\ &\Big/
  \Bigg\{\left[\gamma_1+i \left(\omega -2 \omega _1\right)\right]^3
  \left[-i \gamma _2+\omega -2 \omega _1\right) \left(\gamma _1+2
  \left(\gamma _e+i \left(-\omega _e+\omega -\omega
  _1\right)\right)\right]^2 \left(-2 i \gamma _e-2 \omega _e+\omega
  \right)\Bigg\} \label{geneq}
\end{align}
\end{widetext}
\normalsize Given this unruly but general expression, we consider two
main limiting cases in the main text, corresponding to either a
monochromatic ($\gamma_e \rightarrow 0$) or broadband ($\gamma_e \gg
\gamma_1$) emitter, leading to the equations given in the main
text. For comparison, we also consider second-harmonic generation in
the ND limit, in which case the steady-state amplitude is given by:
\begin{widetext}
\begin{align}
a_2^\text{ND}(\omega)=\frac{16 i \pi \gamma_e \beta^\text{ND}
  s^\text{ND}}{\left(-i \gamma _1+\delta\right) \left(-i \gamma
  _2+\delta\right) \left(\gamma _1+2 \left(\gamma _e+i \left(-\omega
  _e+\omega -\omega _1\right)\right)\right) \left(-2 i \gamma _e-2
  \omega _e+\omega \right)},
\end{align}
\end{widetext}
where $s^\text{ND}={s_a + s_b \over \sqrt{2}}$ and $\beta^\text{ND} =
{\beta_1 + \beta_2 + \beta_3 \over 2}$.

\bibliographystyle{unsrt} 
\bibliography{nlep,zim,ep3}

\end{document}